\documentclass[%
 reprint,
 amsmath,amssymb,
]{revtex4-1}

\usepackage{graphicx}
\usepackage{dcolumn}
\usepackage{bm}
\usepackage{color}
\usepackage{ulem}

\begin{document}

\title{Simple model for the kinetics of stimuli-responsive gels with porous structures}

\author{Yuhei Yamada}
\email{yamada.y.bw@m.titech.ac.jp}
\affiliation{%
  Living Systems Materialogy Research Group, International Research Frontiers Initiative, Tokyo Institute of Technology, 4259 Nagatsuta-cho, Midori-ku, Yokohama 226-8501, Japan
}

\author{Shingo Maeda}%
\altaffiliation[Also at ]{%
  Living Systems Materialogy Research Group, International Research Frontiers Initiative, Tokyo Institute of Technology, 4259, Nagatsuta-cho, Midori-ku, Yokohama 226-8501, Japan
}
\affiliation{%
  Department of Mechanical Engineering, Tokyo Institute of Technology, 2-12-1 Ookayama Meguro-ku Tokyo 152-8550, Japan
}%

\date{\today}

\begin{abstract}
Stimuli-responsive gels are the \textcolor{black}{polymer gels} whose volume is varied according to environmental conditions.
It has been reported that stimuli-responsive gels with an inhomogeneous structure exhibit faster volume changes than gels without such structures.
It is understood as a difference in the transfer dynamics of the solvent, though, there are few models for discussing the effect of inhomogeneity explicitly.
In this paper, we propose a simple model for the kinetics of volume change by introducing inhomogeneity as the probability distribution of a random variable that characterizes the structure.
Under this framework, we demonstrate that inhomogeneity actually increases the rate of volume change.
\end{abstract}

\maketitle


%
Stimuli-responsive gels are \textcolor{black}{polymer gels} that vary in volume depending on environmental conditions such as temperature, pH, irradiation, etc\cite{ahn2008}.
In recent years, they have attracted considerable attention in engineering and biomedical fields as smart materials that adapt to the environment\cite{hines2017,lavrador2021}.
In many applications of stimuli-responsive gels, there is demand for increasing the ratio and rate of volume change.
The basic mechanism of the volume change is explained by the Flory-Huggins theory, which is a statistical model that gives the osmotic pressure of \textcolor{black}{polymer gels} by calculating the free energy of the state while considering the entropy and enthalpy for mixing the polymer and solvent\cite{doi2013}.
The gel volume is determined to balance the osmotic pressure and elastic stress of the polymer network.
In the case of the stimuli-responsive gel, environmental conditions change the interaction between the polymer and solvent and alter the osmotic pressure in the gel.
Then, the gel absorbs or expels the solvent, resulting in a change in volume.
In this mechanism, the rate of volume change is limited by either the \textcolor{black}{transport coefficient} of stimulus (like heat) or the rate of \textcolor{black}{deformation of the polymer networks}.
In most cases, the motion of the polymer network is slower than the transfer of the stimulus; thus, it becomes a rate-limiting step of volume change\cite{gehrke1993}.
The motion of a polymer network is well described by the Tanaka-Fillmore theory, in which the motion of a polymer network follows a diffusion process\cite{tanaka1973,li1990}.
\textcolor{black}{
Although the original Tanaka-Fillmore theory cannot treat general anisotropic deformation, an extended theory that considers the coupling of stress and diffusion is proposed, and this theory has succeeded in explaining many of experimental results\cite{yamaue2004,yamaue2005,doi2009}.
In both theories, the rate of volume change is essentially restricted by the friction between the polymer network and solvent.
}

Many efforts have been devoted to enhancing the rate of volume change through control of the synthesis method and chemical conditions; however, recent studies have revealed another approach that is very effective in increasing the rate; introducing inhomogeneity.
Experimental studies have revealed that when a responsive gel is prepared with a microporous open-celled structure, the solvent can be absorbed or expelled by the convection process, which accelerates the swelling/shrinking kinetics much faster than that by the diffusion process of the polymer network\cite{kabra1992,takata2002,kato2004,gotoh2004,maeda2008,murase2009,seiffert2012}.
It has also been reported that an inhomogeneous structure drastically changes the static properties of \textcolor{black}{polymer gels} like equilibrium swelling ratio and elasticity\cite{lorenzo2015,lopez2019}.
From a theoretical perspective, while great developments in numerical simulation methods enable exploring precise dynamics in the interior of responsive gels\cite{nikolov2018,camerin2018}, theoretical understanding based on a simple model that captures essential features seems to be less progressed.
To understand the role of inhomogeneity in the kinetics, we need a model that explicitly treats inhomogeneity.
In this study, we propose a simple model for the kinetics of responsive gels that introduces inhomogeneity as a probability distribution.
Under this framework, we first discuss the role of the inhomogeneous structure and show that the inhomogeneity actually accelerates the kinetics.
%
%


%
We focus on the uniform deformation of \textcolor{black}{polymer gels}.
The volume change in infinitesimal time variation is then equal to the flow rate through the surface.
Denoting the vgel volume as $V$, the flow rate of solvent into the gel as $Q$, and the time as $t$,
\begin{equation}
    \frac{dV}{dt} = Q. \label{volume_change}
\end{equation}
Our strategy is to model the solvent flow in porous gels via the Hagen-Poiseuille flow and introduce an inhomogeneous structure as the distribution of the flow channel.
In the absorbing and expelling process of the solvent in actual \textcolor{black}{polymer gels}, the solvent flows through tortuous channels.
Here, we approximate the channels as a bundle of capillaries with different radii.
Denoting the radius of capillary $j$ as $\lambda_j$ and the flow rate of that as $q(\lambda_j)$, which is positive for the direction from external to internal of the gel, the total flow rate into the gel is given by
\begin{equation}
    Q = \sum_j q(\lambda_j) . \label{total_flowrate}
\end{equation}
For $q(\lambda_j)$, we assume the Hagen-Poiseuille law, which is plausible for the flow in cylindrical pipes at low Reynolds numberss:
\begin{equation}
    q(\lambda_j) = \gamma \frac{\lambda_j^4}{l} \Delta p , \label{flowrate}
\end{equation}
where $l$ is the length of the channel, $\Delta p$ is the pressure difference between the external and internal of the gel, and $\gamma$ is a constant.
The approximation for $Q$ is similar to that used for modeling flow in solid porous media, which is found to be valid for the cases\cite{wu2007,yu2008}.
%
%
$\Delta p$ is computed from various factors reflecting each mechanism of stimuli responsiveness and is given for individual cases.
Although $l$ and $\Delta p$ can be different for each channel, for simplicity, we assume they are identical here.
Note that $\gamma, \lambda_j$ and $l$ are positive according to their definition, though, $\Delta p$ can be positive or negative.
The \textcolor{black}{sign} of $\Delta p$ determines the direction of flow; thus, we can apply this framework to both swelling and shrinking processes.
For further simplification, we assume that the shape of the flow channel is scaled by the total volume of channels (which equals to the volume of solvent):
\begin{equation}
    \lambda_j = A_j V_s^{\frac{1}{3}},  \quad l = B V_s^{\frac{1}{3}}, \label{scaling}
\end{equation}
where $V_s$ is the solvent volume and $B$ and $A_j$ are positive constants.
Denoting the volume of the polymer as $V_p$, $V= V_p + V_s$.
Here, $A_j$ are random numbers, and we denote the probability density function as $f(A)$, which represents the inhomogeneity of the structure.
\textcolor{black}{
The summation of the volume of each channel equals $V_s$; thus,
\begin{equation}
    V_s = \sum_j \pi \lambda_j^2 l = V_s \left( \pi B \sum_j A_j^2 \right).
\end{equation}
When there are sufficiently large number of channels, the summation is replaced by an integral.
Then, we obtain
\begin{equation}
    \int A^2 f(A)dA = \frac{1}{\pi B} .  \label{normalization}
\end{equation}
}
This becomes the normalization condition of $f(A)$.
Substituting Eqs. (\ref{total_flowrate}), (\ref{flowrate}), (\ref{scaling}) into Eq. (\ref{volume_change}) and replacing the summation with an integral, we obtain
\begin{align}
    \frac{dV}{dt} 
    &=\frac{\gamma}{B}V_s \int \Delta p A^4 f(A) dA \nonumber \\
    &=\frac{\gamma}{B} (V-V_p) \int \Delta p A^4 f(A) dA.
\end{align}
When the volume fraction of polymer in the gel is sufficiently small, (which is often valid for responsive gels,) $V_p$ becomes negligible compared with $V$.
Then, we obtain
\begin{equation}
    \frac{dV}{dt} = \frac{\gamma}{B}V \int \Delta p A^4 f(A) dA. \label{general}
\end{equation}
This becomes the differential equation for $V$ when $\Delta p$ and $f(A)$ are given.
\textcolor{black}{
We now discuss the effect of inhomogeneity on the rate of volume change.
If $\Delta p$ does not depend on $A$, then Eq. (\ref{general}) becomes
\begin{equation}
    \frac{dV}{dt} = \gamma' \{V \cdot \Delta p(V) \}, \label{general_sim}
\end{equation}
where
\begin{equation}
    \gamma' \equiv \frac{\gamma}{B} \int A^4 f(A) dA . \label{model_gamma}
\end{equation}
$\gamma'$ controls the rate of volume change, which depends on the degree of inhomogeneity, $f(A)$.
For example, consider the case where the probability density function is a Gaussian distribution:
\begin{equation}
    f(A) = \frac{C}{\sqrt{2\pi \sigma^2}} \exp{\left( -\frac{(A-\mu)^2}{2\sigma^2} \right)} , \label{gauss}
\end{equation}
where $\mu, \sigma$ and $C$ are constants representing the mean, standard deviation, and normalization constant, respectively.
To ensure that $A_j>0$, we assume $\mu \gg \sigma$ here.
We now compare cases in which all physical parameters $V, B,\Delta p$ and $\mu$, are the same except $\sigma$.
We can compute $C$ using Eq. (\ref{normalization}) as follows:
\begin{equation}
    C = \frac{1}{\pi B} \frac{1}{\mu^2 + \sigma^2} .
\end{equation}
Then, $\gamma'$ is computed from Eq. (\ref{model_gamma}) as
\begin{equation}
    \gamma' = \frac{\gamma}{\pi B^2} \frac{1}{\mu^2 + \sigma^2} (\mu^4 + 6\mu^2 \sigma^2 + 3\sigma^4) .  \label{gamma_prime}
\end{equation}
It can be seen from Eq. (\ref{model_eq}) that the rate of volume change is proportional to $\gamma'$.
Because the variation of $\sigma$ only affects $\gamma'$ in the right-hand side of Eq. (\ref{model_eq}), one can discuss the effect of $\sigma$ through comparing the value of $\gamma'$:
\begin{equation}
    \gamma'(\sigma \neq 0) - \gamma'(\sigma=0) = \frac{\gamma}{\pi B^2} \frac{5\mu^2 \sigma^2 + 3\sigma^4}{\mu^2 + \sigma^2}\geq 0 .  \label{ineq}
\end{equation}
This indicates that the rate of volume change is larger in the inhomogeneous case ($\sigma \neq 0$) than in the homogeneous case ($\sigma = 0$).
Note that the parallel discussion is valid when $f(A)$ is not a Gaussian but a symmetric distribution around the mean value. 
}

\begin{figure}
\begin{center}
\resizebox{0.4\textwidth}{!}{%
  \includegraphics{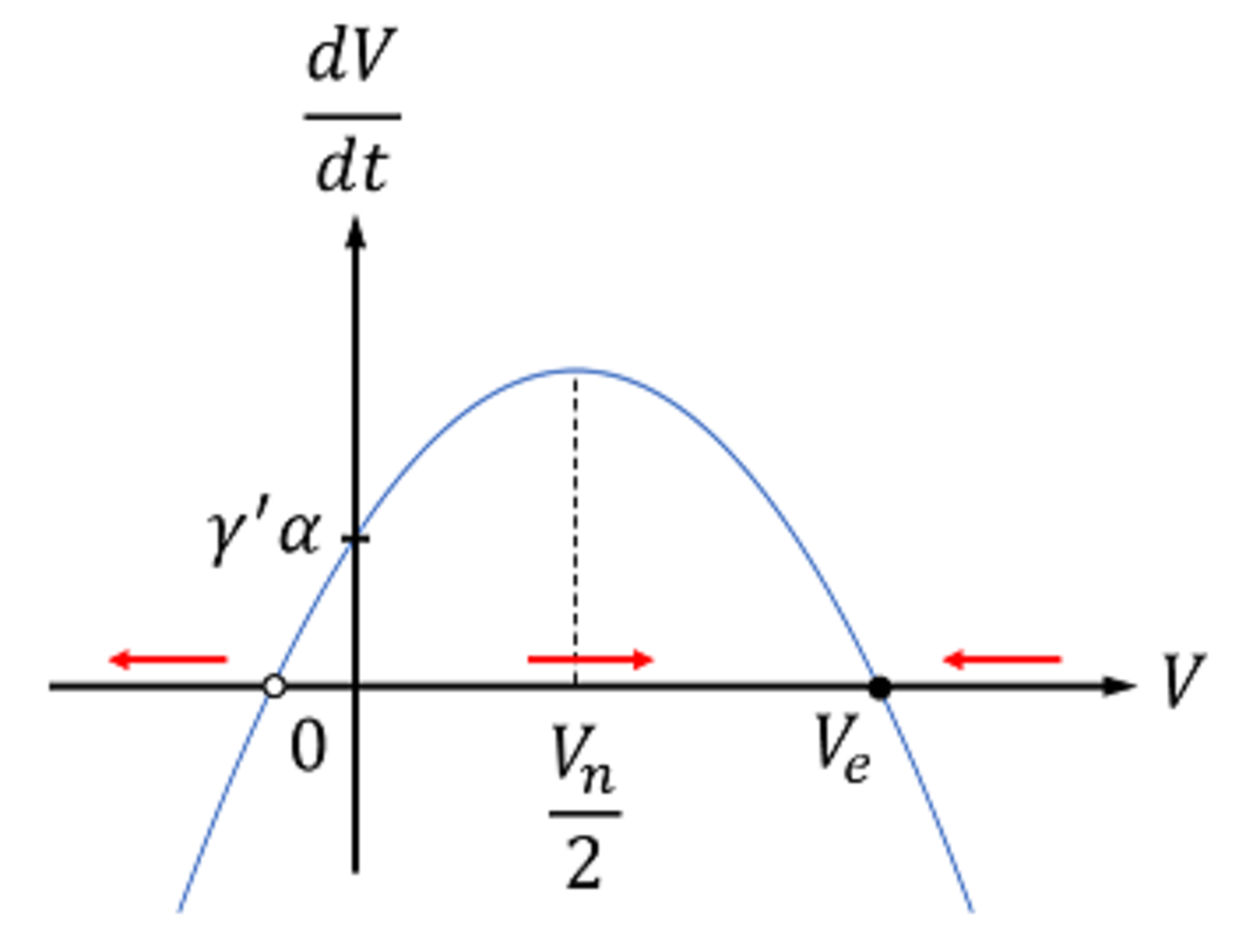}
}
\caption{
Phase space of the differential equation (\ref{model_eq}).
Red arrows show the direction of vector field.
}
\label{phase_space}
\end{center}
\end{figure}

\textcolor{black}{
Now we demonstrate an example with giving specific $\Delta p$.
In actual problems, $\Delta p$ would be a complex function that should be given by a theory or phenomenological way.
Here, we consider a toy model that $\Delta p$ include osmotic pressure-like effect and elastic stress-like effect:
\begin{equation}
    \Delta p (V) = \frac{\alpha}{V} - \beta \left(\frac{V-V_n}{V_n} \right), \label{pressure}
\end{equation}
where $\alpha, \beta$ and $V_n$ are constants.
On the right-hand side, the first term represents the "osmotic pressure", which is proportional to the polymer concentration and so proportional to $V^{-1}$.
And the second term represents the "elastic stress", which is proportional to the volume strain, $(V-V_n)/V_n$.
$\alpha$ and $\beta$ control the strength of the effects, and $V_n$ represents "natural volume" that elastic stress becomes zero when $V=V_n$.
}
To exclude physically meaningless cases, we assume $\beta>0$ and $V_n>0$.
By substituting Eq. (\ref{pressure}) into Eq. (\ref{general_sim}), we obtain
\begin{equation}
    \frac{dV}{dt} = \gamma' \left\{\alpha - \beta\frac{V}{V_n} \left(V-V_n\right) \right\} , \label{model_eq}
\end{equation}
This equation is equivalent to the logistic equation.
Although there are four parameters in Eq. (\ref{model_eq}), since $\alpha, \beta$ and $\gamma'$ attribute to the equation only in the form of $\alpha$ and $\gamma' \beta$, the number of relevant parameters is three.
Figure \ref{phase_space} shows a schematic of the phase space of Eq. (\ref{model_eq}).
$V_e$ indicates the equilibrium volume corresponding to a fixed point in the dynamics.
Requiring the existence of a stable fixed point, $\alpha > -\beta V_n/4$ is assumed.
When $\alpha<0$, there is a solution that $V \to -\infty$.
This is a physically nonsense solution due to the simplicity of the assumption for $\Delta p$.
Thus, we ignore such an initial condition in the following discussion.

\begin{figure}
\begin{center}
\resizebox{0.45\textwidth}{!}{%
  \includegraphics{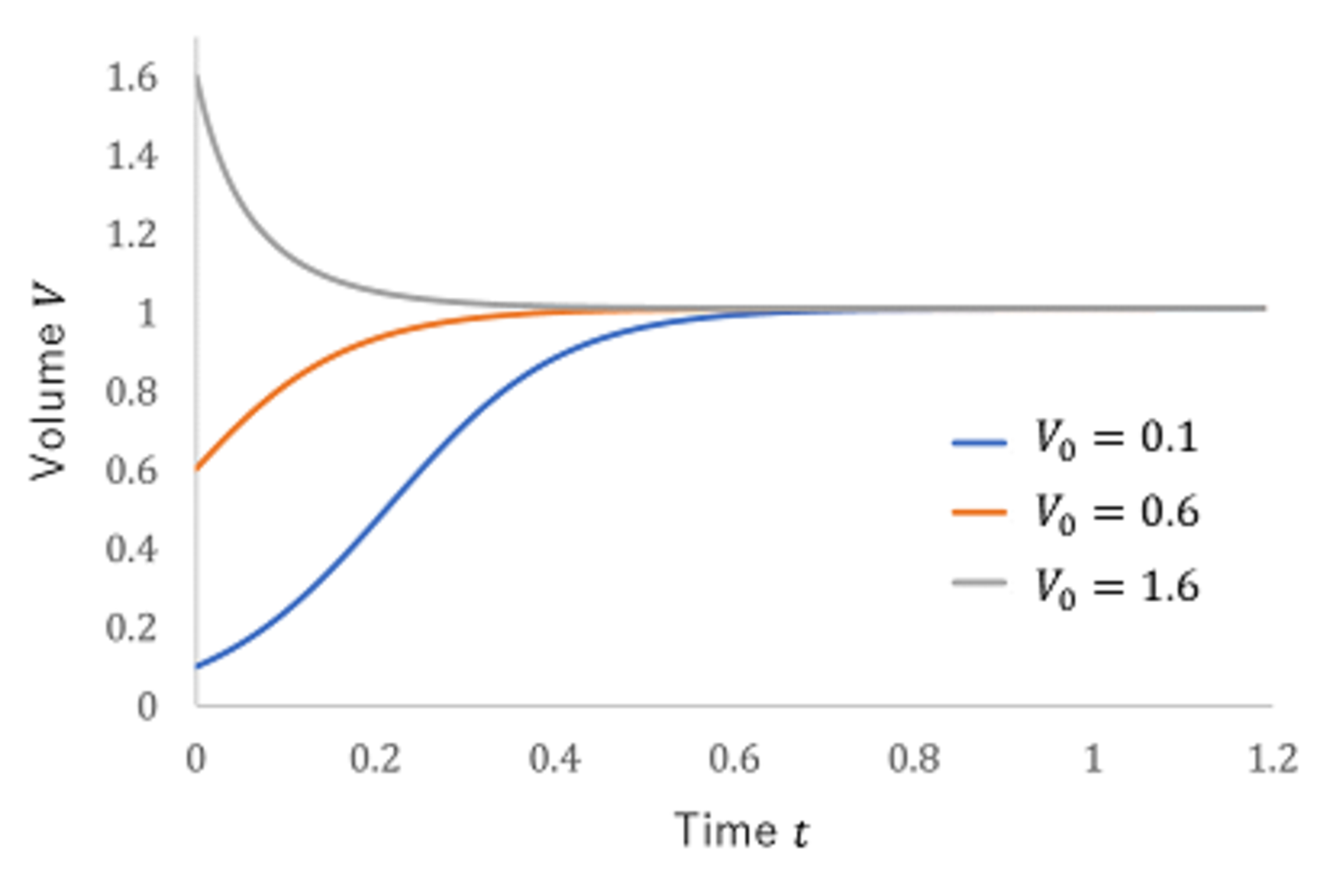}
}
\caption{
Typical time evolution of the volume in the model described by Eq. (\ref{model_eq}).
($\alpha = 0.1, \beta = 10, \gamma'=1, V_n = 1$ and $V_0=0.1, 0.6, 1,6$.)
}
\label{evolution}
\end{center}
\end{figure}

Under the initial condition ($V(0)=V_0$), we can integrate differential equation (\ref{model_eq}) as follows:
\begin{equation}
    V(t) = \frac{-a}{c e^{t/\tau}+1} + V_e , \label{solution}
\end{equation}
where 
\begin{align}
\begin{split}
    a = V_n \left( 1 + \frac{4\alpha}{\beta V_n}\right)^{\frac{1}{2}} , \quad
    \tau = \frac{1}{\gamma' \beta} \left( 1 + \frac{4\alpha}{\beta V_n}\right)^{-\frac{1}{2}} \\
    V_e = V_n \left\{ \frac{1}{2} + \frac{1}{2}\left( 1 + \frac{4\alpha}{\beta V_n}\right)^{\frac{1}{2}} \right\}, \quad
    c = \frac{V_e+V_0-V_n}{V_e-V_0}.
\end{split} \label{solution_parameter}
\end{align}
The four parameters described by Eqs. (\ref{solution_parameter}) correspond to the set of the three relevant parameters in Eq. (\ref{model_eq}) and the initial value of $V$.
Figure \ref{evolution} shows three typical behaviors of the proposed model.
Parameters are set as $\alpha = 0.1, \beta = 10, \gamma'=1, V_n = 1$, and the cases for the three initial conditions $V_0=0.1, 0.6, 1.6$, are presented.
\textcolor{black}{
Although Eq. (\ref{solution}) is derived from a rough assumption for $\Delta p$ as Eq. (\ref{pressure}), it mimics some features of actual behavior like exponential relaxation or the sigmoid curve\cite{li1990,suzuki2001,kaneko1995}.
}

\begin{figure}
\begin{center}
\resizebox{0.45\textwidth}{!}{%
  \includegraphics{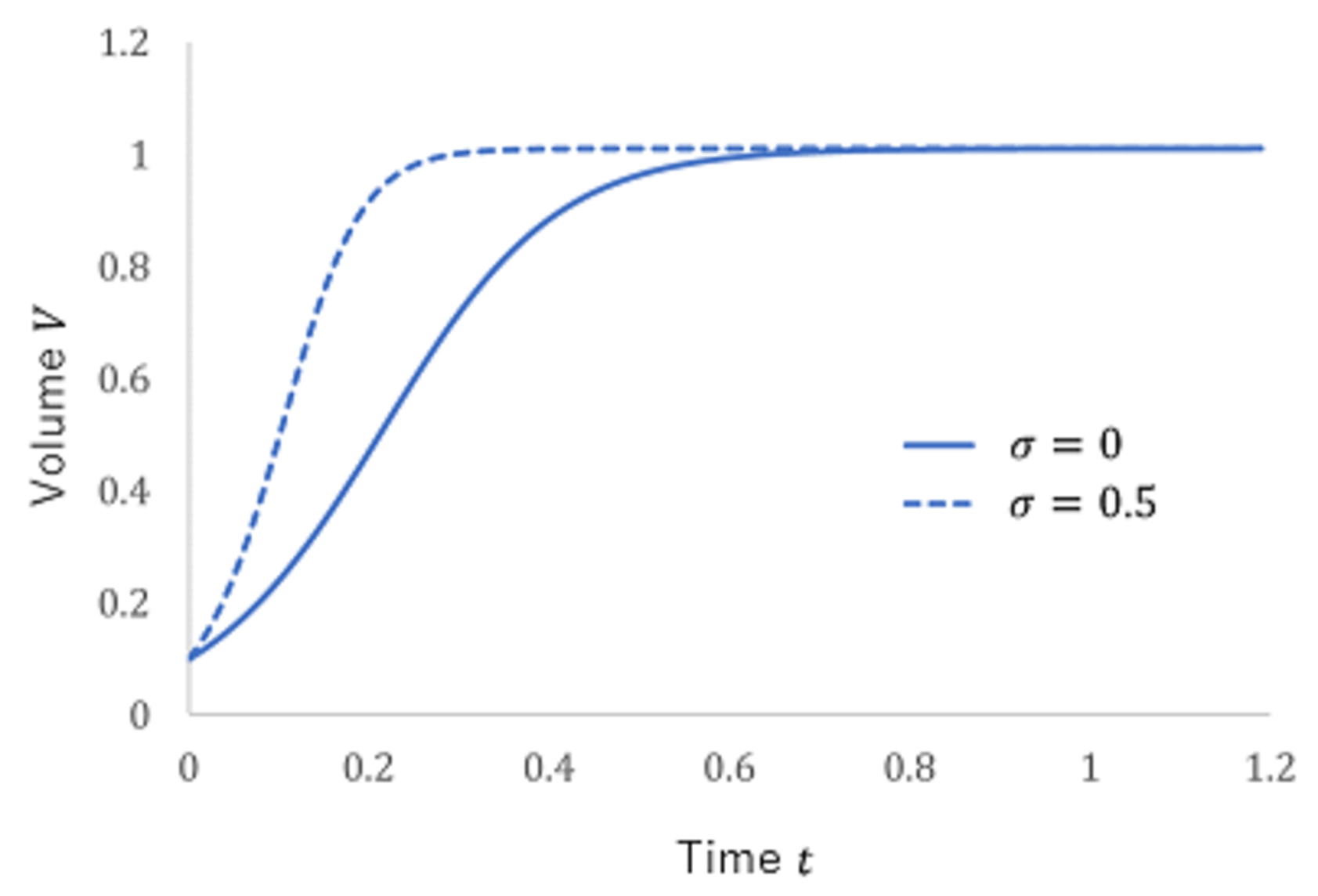}
}
\caption{
Comparison of the time evolution of volume obtained by the model described by Eq. (\ref{model_eq}) for different $\sigma$.
$\gamma'$ is computed by Eq. (\ref{gamma_prime}).
Curve and dashed curve correspond to the cases for $\sigma=0$ and 0.5, respectively. 
($\alpha = 0.1, \beta = 10, V_n = 1, V_0=0.1, \gamma/\pi B^2=1$ and $\mu=1$.)
}
\label{comparison}
\end{center}
\end{figure}

\textcolor{black}{
Under this model, we check the effect of inhomogeneity.
Again, we adopt a Gaussian distribution for $f(A)$ as Eq. (\ref{gauss}).
Fixing $\gamma, B, \mu$ and $\sigma$, we can compute $\gamma'$ by Eq. (\ref{gamma_prime}).
For example, when $\gamma/\pi B^2=1$ and $\mu=1$, $\gamma'=1$ if $\sigma=0$ and $\gamma'=5$ if $\sigma=0$.
Figure \ref{comparison} shows the time evolution of the two cases of $\sigma$ with fixing other parameters.
As discussed by using inequality (\ref{ineq}), it can be seen from the figure that volume change becomes faster for larger $\sigma$.
}

In summary, we propose a simple model for the volume change of stimuli-responsive gels with a porous structure.
We approximated the structure of the porous gel using a bundle of capillaries with different radii and modeled the rate of volume change as equal to the hydrodynamic flow rate.
In the framework, the driving force is the pressure difference, $\Delta p$, between the internal and external of the gel, which should be given for individual cases.
For example, we discuss a simple case when $\Delta p$ consists of the "osmotic pressure" and "elastic stress".
From the analysis, we confirmed that the inhomogeneity of the structure actually increases the rate of volume change.
\textcolor{black}{
We believe that our model helps understanding of kinetics of inhomogeneous stimuli-responsive gels and also contributes to the design of gels in soft matter engineering.
}

\bibliography{apssamp}

\end{document}